\documentclass[12pt]{article}

\usepackage{scicite}

\input epsf

\newenvironment{sciabstract}{%
\begin{quote} \bf}
{\end{quote}}

\newcounter{lastnote}
\newenvironment{scilastnote}{%
\setcounter{lastnote}{\value{enumiv}}%
\addtocounter{lastnote}{+1}%
\begin{list}%
{\arabic{lastnote}.}
{\setlength{\leftmargin}{.22in}}
{\setlength{\labelsep}{.5em}}}
{\end{list}}

\title{A Cyclic Model of the Universe}

\author
{Paul J. Steinhardt$^{1\ast}$ and Neil Turok$^{2}$\\
\\
\normalsize{$^1$Joseph Henry Laboratories,
Princeton University, Princeton, NJ 08544, USAA}\\ 
\normalsize{$^{2}$DAMTP, CMS, Wilberforce Road, Cambridge, CB3 0WA, UK}\\
\\
\normalsize{$^{\ast}$To whom correspondence should be addressed; E-mail:  steinh@princeton.edu.}
}

\date{}

\begin{document} 


\baselineskip24pt


\maketitle


\begin{sciabstract}
We propose a cosmological model in which
the universe undergoes an endless sequence
of cosmic epochs each beginning  with  a `bang'
and ending  in a `crunch.'  The temperature and density 
are finite  at each transition from crunch to  bang.
Instead of having an inflationary epoch,
each cycle includes a period  of slow
accelerated expansion (as recently observed) followed by 
slow contraction. The combination
 produces
the  homogeneity, flatness,
density fluctuations and energy needed to begin the
next cycle.
\end{sciabstract}


\section*{Introduction}
The current standard model of cosmology combines the original big 
bang model  and the inflationary 
scenario.\cite{Gut,Lin,Lin2,BST,BST2,BST3,BST4}
Inflation, a  brief    
period ($10^{-30}$~s) of very rapid cosmic acceleration occurring
shortly
after the big bang, can explain 
 the homogeneity and isotropy  of the universe on large scales ($>100$~Mpc),
its spatial flatness,  and also the distribution of galaxies and 
the  fluctuations in the cosmic microwave background.
However, the standard model has some cracks and gaps.
The recent discoveries of  cosmic acceleration  indicating
self-repulsive dark energy\cite{supernova,supernova2,supernova3,rev}  
were  not predicted and 
have no clear
role in the standard model.\cite{Gut,Lin,Lin2}
Furthermore, no explanation is offered for the `beginning of 
time', the initial conditions of the universe, 
or  the long-term future.

In this paper, we present a new cosmology consisting of
an endless sequence of cycles of expansion and contraction.
By definition, there is neither a beginning  nor an
end of time, nor a
need to specify initial conditions.
We  explain the role of dark energy,
and generate the homogeneity, flatness, and density
fluctuations without invoking inflation.

The model we present  is reminiscent 
of  oscillatory models introduced in the 1930's 
based on a closed universe that 
undergoes a sequence of expansions, contractions and bounces. 
The oscillatory models had the difficulty
of having to pass through a singularity in which the energy and 
temperature diverge.  Furthermore, as pointed out by Tolman \cite{Tolman,DP},
entropy produced during one cycle would add to the entropy produced in 
the next, causing each cycle to be longer than the one before it.  
Extrapolating backward in time, the universe would have to have originated
at some finite time in the past so that  the problem of explaining 
the `beginning of time' remains.   
Furthermore, recent measurements of the cosmic microwave background 
anisotropy and large scale structure favor a flat universe over
a closed one.

In our cyclic model, the universe is infinite and flat, rather than 
finite and closed.
We introduce a 
negative potential energy rather than spatial curvature 
to cause  the reversal from expansion to contraction.  Before  
reversal, though, 
the  universe undergoes the usual period of radiation and matter
domination, followed by a long period 
of accelerated expansion 
 (presumably the acceleration that 
has been recently detected\cite{supernova,rev}).
The accelerated expansion, caused by dark energy,
is an essential feature of our model, needed to dilute 
the entropy, black holes and other debris 
produced in the previous cycle so that the universe
is returned to its original pristine vacuum state
before it begins to contract, 
bounce, and begin a cycle anew.

\section*{Essential Ingredients}

As in inflationary cosmology, the cyclic scenario can be 
described in terms of the evolution of a scalar field $\phi$ in 
a potential $V(\phi)$ according to  a  conventional
four-dimensional quantum field theory.
The essential differences is in  the form of the potential and 
the couplings between the scalar field, matter and 
radiation.

The analysis of the cyclic model follows from the action $S$
describing  gravity,
the scalar field $\phi$, and the matter and radiation fluids:
\begin{equation}
S=\int d^4x \sqrt{-g}\left(\frac{1}{16 \pi G}{\cal R}-
\frac{1}{2}(\partial\phi)^2-V(\phi) + \beta^4(\phi)(\rho_M + \rho_R) \right),
\label{eq:4d}
\end{equation}
where $g$ is the determinant of the metric $g_{\mu \nu}$, $G$ is Newton's
constant and ${\cal R}$ is the Ricci scalar. 
The coupling  $\beta(\phi)$ between 
$\phi$ and the matter ($\rho_M$) and radiation ($\rho_R$) densities 
is crucial because it allows the densities to remain finite at the big
crunch/big bang transition.

The line element for a flat, homogeneous universe is 
$-d t^2 + a^2 d {\bf x}^2$, where $a$ is the Robertson-Walker 
scale factor.
The equations of motion following from Eq.~(\ref{eq:4d}) are,
\begin{equation}
\label{eq1}
H^2  =  \frac{8 \pi G}{3}
\left( \frac{1}{2} \dot{\phi}^2 +V + \beta^4 \rho_R + 
\beta^4 \rho_M \right),
\end{equation}
\begin{equation}
\frac{\ddot{a}}{a}  =  - \frac{8 \pi G}{3} \left(\dot{\phi}^2 -V + 
\beta^4 \rho_R
+{1\over 2} \beta^4 \rho_M \right),
\label{eq2}
\end{equation}
where  a  dot denotes a  derivative with respect to $t$ and 
 $H \equiv \dot{a}/{a}$ is the Hubble parameter.
The equation of motion for $\phi$ is
\begin{equation}
\label{eq2a}
\ddot{\phi}+3 H \dot{\phi} =  -V_{,\phi} - \beta_{,\phi} \beta^3 \rho_M
\end{equation}
and the fluid equation of motion for matter (M) or radiation (R) is
\begin{equation}
\hat{a} \frac{d \rho_i}{d \hat{a}} = a \frac{\partial \rho_i}{\partial a}
+ \frac{\beta}{\beta'} \frac{\partial \rho_i} {\partial \phi_i} = 
-3 (\rho_i + p_i), \, \,\qquad i = {\rm M}, {\rm R} \, \, ,
\end{equation}
where $\hat{a} \equiv a \beta(\phi)$ and
$p$ is the pressure of the fluid component with energy density 
$\rho$. The implicit assumption is that matter and radiation couple to 
$\beta^2(\phi) g_{\mu \nu}$ (with scale factor $\hat{a}$)
 rather than the 
Einstein metric $g_{\mu \nu}$ alone (or the scale factor $a$).
Note that the radiation term in Eq.~(\ref{eq:4d}) 
is actually independent of $\phi$ (since $\rho_R \propto \hat{a}^{-4}$)
so only $\rho_M$ enters the $\phi$ equation of motion. 

\begin{figure}
\begin{center}
\epsfxsize=3.5 in \centerline{\epsfbox{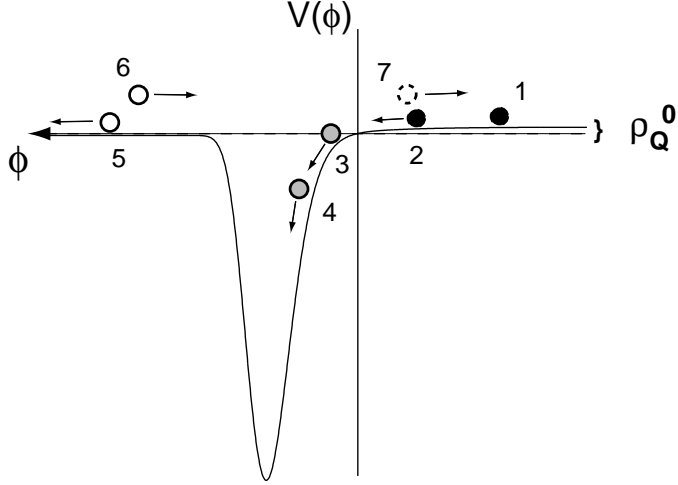}}
  \end{center}
 \caption{
Schematic plot of the potential $V(\phi)$ versus
field $\phi$.  In M theory, $\phi$ determines
the distance between branes, and  $\phi \rightarrow -\infty$
as the branes collide. We define $\phi$ to be zero where $V(\phi)$
crosses zero and, therefore, $\phi$ is positive when the branes
are at their maximal separation.  Far to the right, the potential
asymptotes to $\rho_Q^0$, the current value of the quintessence (dark
energy) density. The solid circles represent the dark energy dominated
stage; the grey represent the contracting phase during which 
density fluctuations are generated; the open circles represent 
the phase when the scalar kinetic energy dominates; and the broken 
circle represents the stage when the universe is radiation dominated.
Further details of the  sequence of
stages are described in the article. 
} \end{figure}

We assume the potential $V(\phi)$ has the following 
features,  illustrated in  Fig.~1:
($i$) 
$V(\phi)$ must approach zero rapidly as $\phi \rightarrow -\infty$;
($ii$)  the potential must be negative for intermediate $\phi$;
and, ($iii$)
 as $\phi$ increases, the potential must rise  to a shallow 
plateau with a positive value $V_0$.  An example of a potential 
with these properties is
\begin{equation} \label{potential}
V(\phi) = V_0 (1- e^{-c \phi}) F(\phi),
\end{equation} 
where  from this point onwards we adopt units 
in which $8 \pi G=1$.
$F(\phi)$ is a function we introduce to ensure that $V(\phi) \rightarrow 0$
as $\phi \rightarrow -\infty$.  
Without loss of generality, we take 
$F(\phi)$ to be nearly unity
for $\phi$ to the right of 
potential minimum. The detailed manner in which it tends to
zero  is not 
crucial for the main 
predictions of the cyclic model.   A  quantitative analysis
of this  potential (Ref.~\ref{other}) shows that a realistic
cosmology can be obtained by choosing $c\ge 10$ and
$V_0$  equal to 
today's dark energy density  (about $6 \times 10^{-30}$~g/cm$^3$)
in Eq.~(\ref{potential}).

We have already mentioned that the coupling $\beta(\phi)$ is chosen
so that  $\hat{a}$ and, thus, the matter and radiation density
are finite at $a=0$.  This requires
$\beta(\phi) \sim {\rm e}^{-\phi\sqrt{6}}$
as $\phi \rightarrow -\infty$, but this is precisely the behavior
expected in M-theory (see below).  The presence of $\beta(\phi)$ and
the consequent coupling of $\phi$ to nonrelativistic matter 
represent a modification  of Einstein's theory of general relativity.
Because   $\phi$ evolves 
by an exponentially small  amount between nucleosynthesis ($t \sim 
1$~s) and today ($t\sim 10^{17}$~s),
deviations from standard general relativity are small and
 easily satisfy
 current cosmological constraints.\cite{other}
However, the coupling of matter to $\phi$
produces other  potentially measurable effects including
 a fifth force which violates the equivalence principle.
Provided $({\rm ln} \beta)_{,\phi} \ll 10^{-3}$, for today's value
of $\phi$, these violations are
 too small to be detected.\cite{other,poly2,poly} We shall assume this
to be the case. 
Hence, the deviations from general relativity are negligible today.

The final crucial ingredient in the cyclic model is a matching 
rule which determines how to pass from the big crunch to the 
big bang.  The transition occurs as $\phi \rightarrow -\infty$
and then rebounds towards positive $\phi$.   
Motivated again by string theory (see below), we propose that some 
small fraction of the $\phi$-field kinetic energy is converted
to matter and radiation.
The matching rule amounts to
\begin{equation} \label{collide}
\dot{\phi} \, {\rm e}^{\sqrt{3/2} \phi} \rightarrow
- (1 + \chi)  \dot{\phi} \,{\rm e}^{\sqrt{3/2} \phi}
\end{equation} 
where  $\chi$ is a parameter measuring the efficiency 
of production of radiation at the bounce.
Both sides of this relation are finite at the bounce.

\section*{Stringy motivation}

From the perspective of four-dimensional quantum field theory,
the introduction of a scalar field, a potential
and the couplings to matter in the cyclic model
is  no more arbitrary or tuned than the requirements for
the inflationary models.  
However, the ingredients for the cyclic model are also 
strongly motivated  by string theory and $M$-theory. 
This connection 
 ties our scenario into the leading approach to
fundamental physics and  quantum gravity.  
The connection should not be overemphasized.
String theory remains far from  proven  and
quantum gravity effects may be
unimportant for describing cosmology 
at wavelengths much longer than a Planck length ($10^{-33}$~cm).
If the reader prefers, the connection to string theory
can be ignored. On the other hand, we find the connection useful
because it provides a natural geometric interpretation for the scenario.
Hence, we briefly describe the relationship.

According to $M$-theory,
the universe
 consists of a four dimensional
 `bulk' space bounded by two three-dimensional  domain walls,
 known as `branes' (short for membranes), 
 one with positive and the other with negative
 tension.\cite{REFBRANE,HW,polch} The
 branes are free to move along the extra spatial
 dimension, so that they may approach and collide.
 The fundamental theory  is formulated in ten spatial dimensions, but
 six dimensions are compactified on a Calabi-Yau
 manifold, which for our purposes can be treated as fixed, and therefore
 ignored.
 Gravity acts throughout the five dimensional spacetime, but
 particles of our visible universe are constrained
 to move along one of the branes, sometimes called the
visible brane.   Particles on the other brane interact
 only through gravity with matter on the visible brane and hence behave like
 dark matter.

The scalar field $\phi$ we want is naturally identified with the
field that determines the distance between branes.  The potential
$V(\phi)$ is the inter-brane potential caused by non-perturbative
virtual exchange of membranes between the boundaries.  
The interbrane force is what causes the branes to repeatedly
collide and bounce.
At large separation (corresponding to large $\phi$), 
the force between the branes should become small, consistent with
the flat plateau shown in Fig.~1.  Collision 
corresponds to $\phi \rightarrow -\infty$. 
But the string coupling 
$g_s \propto e^{\gamma \phi}$,  with $\gamma > 0$, so $g_s$ vanishes
 in this limit \cite{nonsing}. Non-perturbative effects vanish
faster than any power of $g_s$, for example as 
$e^{-1/g_s^2} $   or $e^{-1/g_s}$, accounting for the
prefactor $F(\phi)$ in 
Eq.~(\ref{potential}).

The coupling  $\beta(\phi)$ also
has a natural interpretation in the brane picture.  
Particles reside on the branes, which are embedded in an
extra dimension whose size and warp are determined by 
$\beta$.  The effective scale factor on the branes is
$\hat{a} = a \, \beta(\phi)$, not $a$, and
$\hat{a}$ is finite at the big crunch or big bang.  The function 
$\beta(\phi)$ is in general different for the two branes (due to the 
warp factor) and for different reductions of M-theory.  However,
the standard Kaluza-Klein  behavior 
$\beta(\phi) \sim {\rm e}^{-\phi/\sqrt{6}}$
as $\phi \rightarrow - \infty$ is universal, since
 the warp factor becomes irrelevant as the branes approach one 
 another.\cite{other,nonsing}

Most importantly, the brane-world provides a natural resolution
of the cosmic singularity.\cite{other,nonsing}
One might say that
the big crunch is an illusion, because the scale factors
on the branes ($\rightarrow \hat{a}$)  are perfectly finite there. 
That is why the matter and radiation densities, 
and the Riemannian curvature on the branes, are finite.  
The only respect in which the 
big crunch is singular is that 
the one extra dimension separating the two branes 
momentarily disappears. 
Our scenario is built on the hypothesis\cite{kost} that 
the branes separate after collision, so the extra dimension
immediately reappears. This process cannot be completely
smooth, because the disappearance of the extra dimension
is non-adiabatic and leads to particle production. That is,
the brane collision is partially inelastic. Preliminary
calculations of this effect are encouraging, because they
indicate a finite density of particles is produced \cite{newpolch,newpolch2}. 
The matching condition, Eq.~(\ref{collide}), parameterizes this effect.
Ultimately, a well-controlled string-theoretic 
calculation\cite{other,newpolch,newpolch2,nonsing}
should determine the value of $\chi$.

\section*{Dark energy and the cyclic model}

The role of dark energy in the cyclic scenario is novel.
In the standard big bang and inflationary models, the recently 
discovered dark energy and cosmic acceleration\cite{supernova,rev}
are an unexpected
surprise with no clear  explanation.
In the cyclic scenario, however,
not only is the source of dark energy explained, but the 
dark energy and its 
associated cosmic acceleration 
are actually crucial to the consistency of the model.
Namely, the associated exponential expansion suppresses 
density perturbations and 
dilutes entropy, matter and black holes to negligible 
levels.
By periodically 
restoring the universe to 
an empty, smooth state, the acceleration 
causes the cyclic solution to be  a stable attractor.

Right after a big bang, the scalar field $\phi$ is 
increasing rapidly. However, its motion is damped 
by the expansion of the universe and $\phi$ essentially
comes to rest in the 
radiation dominated phase
[stage (1) in Figure 1]. Thereafter it remains nearly fixed
until 
the dark energy begins to dominate and cosmic acceleration commences.
The positive potential energy density at the current value
of $\phi$  acts as a form of quintessence,\cite{quint}
a time-varying energy component with negative pressure that
causes the present-day accelerated expansion.
This choice entails tuning $V_0$, 
but it is the same degree of  tuning required in any
cosmological model (including inflation) to explain the 
recent observations of cosmic acceleration \cite{supernova,rev}.  
In this case, because the dark energy serves several purposes,
the single tuning  resolves several problems at once.

The cosmic acceleration is nearly 100 orders of magnitude smaller
than considered in inflationary cosmology.
Nevertheless, if sustained for hundreds of e-folds 
(trillions of years) or more,
the cosmic acceleration
can flatten the universe and dilute
the entropy, black holes, and other debris (neutron stars, neutrinos, {\it etc.})
created over the preceding
cycle, overcoming the obstacle that has blocked previous attempts
at a cyclic  universe. In this picture, we are presently about 14 billion
years into the current cycle, and have just begun the trillions years
of cosmic acceleration. After this amount of accelerated expansion,
the number of particles in the universe 
may  be
suppressed to less than one per Hubble volume
before the cosmic acceleration ends.
Ultimately, the scalar field begins to roll back
towards $-\infty$, driving the potential to zero.
The scalar field $\phi$ 
is thus the source of the currently observed acceleration, the reason
why the universe is homogeneous, isotropic and flat before the
big crunch, and the
root cause for the  universe reversing from expansion to contraction.

\section*{A brief tour of the cyclic universe}

Putting together the various concepts that have been introduced,
we can now present 
the sequence of events in each cycle 
beginning from the present epoch, 
stage (1)  in Figure 1.
The universe has completed radiation and matter dominated epochs
during which $\phi$ is nearly fixed.
We are presently at the time when its 
potential energy  begins 
to dominate, ushering in 
a period of slow cosmic acceleration
lasting trillions of years or more, in which the
matter, radiation and black holes are diluted away
and a smooth, empty, flat universe results.
 Very slowly the slope
in the potential causes 
$\phi$ to roll in the negative direction, as
indicated in stage (2).  
Cosmic acceleration continues until the field  nears the point of
zero potential energy, stage (3).  The universe is dominated by the 
kinetic energy of $\phi$, but expansion causes this to be damped.
Eventually, 
the total energy (kinetic plus negative potential) reaches zero. 
From 
Eq.~(\ref{eq1}),  the
Hubble parameter is zero and the universe is momentarily static.
From Eq.~(\ref{eq2}), $\ddot{a} <0$, so that $a$
begins to contract. While $a$ is nearly static,
the universe satisfies the ekpyrotic conditions for
creating  a scale-invariant spectrum of 
density perturbations.\cite{kost,ekperts}
As the field continues to roll towards $-\infty$,
 $a$ contracts and the 
kinetic energy of the scalar field grows.
That is, gravitational energy is converted to
scalar field kinetic energy during this part of the cycle.
Hence, the field races past the minimum of the potential and
off to 
$-\infty$, with 
kinetic energy becoming increasingly dominant as the bounce
nears, stage (5). The scalar field diverges as 
$a$ tends to zero.
After the bounce, radiation is generated and 
the universe is expanding.
At first, scalar kinetic energy density 
($\propto 1/a^6$) dominates over the radiation ($\propto 1/a^4$), 
 stage (6).  Soon after, however, 
 the universe becomes radiation dominated, stage (7).
The motion of $\phi$ is rapidly damped away, so that it
remains close to its maximal value for the rest of the 
standard big bang evolution 
(the next 15 billion  years).
Then, the scalar field potential energy  begins to dominate,
and the field rolls towards $-\infty$, where the next
big crunch occurs and the cycle
begins anew.

\section*{Obtaining scale-invariant perturbations}

One of the most compelling successes of inflationary theory was to
obtain a nearly
scale-invariant spectrum of density 
fluctuations that can seed large-scale
structure.\cite{BST}
Here, the same feat is achieved using 
different physics
during an ultra-slow contraction phase 
[stage (2) in Fig.~1].\cite{kost,ekperts}
In inflation, the  density fluctuations are 
created by very rapid expansion, causing
fluctuations on microscopic
scales to be stretched to 
macroscopic scales.\cite{BST}
In the cyclic model,
the fluctuations are generated during
a quasistatic, contracting universe where
gravity plays no significant role.\cite{kost}  
Simply because the potential $V(\phi)$
is decreasing more and more rapidly, 
quantum fluctuations in $\phi$ are amplified
as the field  evolves downhill.\cite{kost,linde,reply}   
Instabilities in long-wavelength modes occur sooner than those in 
short wavelength modes, thereby amplifying long wavelength power
and, curiously, 
nearly exactly mimicking the inflationary effect.
The nearly scale-invariant
spectrum of fluctuations in $\phi$ 
created during
the contracting phase transform into a nearly 
scale-invariant spectrum of density fluctuations
in the expanding phase.\cite{ekperts}  
Current observations of large-scale structure and fluctuations
of the cosmic microwave background cannot distinguish between 
inflation and the cyclic  model because both predict a 
nearly scale-invariant spectrum of adiabatic, gaussian 
density perturbations.

Future measurements of 
gravitational waves may be able to distinguish the two pictures.\cite{kost}
 In inflation, where gravity is paramount,
quantum fluctuations 
in all light degrees of freedom are subject
to the same gravitational effect described above. Hence, not only 
is there a nearly scale-invariant spectrum of energy density 
perturbations, but also there is a scale-invariant spectrum of 
gravitational waves.  
In the cyclic and ekpyrotic models, where the potential, rather than 
gravity, is the cause of the fluctuations, the only field which obtains
a nearly scale-invariant spectrum is the one rolling down the 
potential, namely $\phi$, which only produces energy density 
fluctuations.  The direct search for gravitational waves 
or the search for their indirect effect on the polarization of the 
cosmic microwave background\cite{grav} 
are the crucial tests for distinguishing
inflation from the cyclic model. 

\section*{Cyclic solution as Cosmic Attractor}

Not only do 
cyclic solutions exist for a range of 
potentials and parameters, but also  they are 
attractors for a range  of initial conditions.  
The cosmic acceleration caused by the positive 
potential plateau 
plays the critical role here.
For example,
suppose the scalar field is jostled and stops 
at a slightly different maximal value on the plateau
compared to the exactly cyclic solution. The same sequence 
of stages ensues.
The scalar field is critically
damped during the exponentially expanding phase.
So by the time
the field reaches stage (3) where $V=0$, it is rolling almost
at the 
same rate as if it had started at $\phi=0$, and 
memory of its initial 
position has been lost.\cite{other}
The argument suggests that it is natural to expect 
dark energy and cosmic acceleration
following matter domination in a cyclic universe, in accordance
with what has been recently observed.

\section*{Comparing  cyclic and inflationary  model}

The cyclic and inflationary models have numerous conceptual 
differences in addition to those already described.
Inflation requires two periods of cosmic acceleration, a
hypothetical period of rapid expansion in the early universe 
and the observed current acceleration.  The cyclic model only
requires one period  of acceleration per cycle.

In the inflationary picture, most of the volume of the universe is
completely unlike what we see.  Even when inflation ends in one region,
such as our own,
it continues in others.  Because of the superluminal expansion 
rate of the remaining inflating regions, they occupy most of the 
physical volume of the universe.  Regions which have stopped
inflating, such as our 
region of the universe, represent an infinitesimal fraction.
By contrast, the cyclic model is one in which the local universe is
typical of the universe as a whole.  All or almost all regions of the
universe are undergoing the same sequence of cosmic events and most
of the time is spent in the radiation, matter, and dark energy dominated
phases.  

In the production of perturbations, the inflationary mechanism
relies on stretching modes 
whose wavelength is initially 
exponentially sub-Planckian,
to macroscopic scales. Quantum gravity effects in the initial state
are highly  uncertain, and inflationary predictions may therefore be
highly sensitive to sub-Planckian physics. In contrast, 
perturbations in the cyclic model 
are generated when the modes have wavelengths of thousands of
kilometers, using macroscopic physics insensitive to quantum gravity
effects. 

The cyclic model deals directly with the cosmic singularity,  
explaining it as a transition from a 
contracting to an expanding phase.  
 Although
inflation does not address the cosmic singularity problem directly,
 it does rely implicitly  on the opposite assumption:
 that the big bang is the beginning of time and that
the universe emerges in a rapidly expanding state.
Inflating regions with high 
potential energy expand more rapidly and dominate the universe.
If there is a pre-existing contracting phase, then 
the high potential energy regions collapse and disappear
before the expansion phase begins.  String theory or, more generally,
quantum gravity can play an important role in settling the nature of
the singularity and, thereby, distinguishing between
the two assumptions.

The cyclic model is a complete model of cosmic history,
whereas inflation is only a theory of cosmic history following
an assumed initial creation event.
Hence, the cyclic model has more  
explanatory and  predictive power.
For example, we have already emphasized how the cyclic
model leads naturally to the prediction of quintessence
and cosmic acceleration, explaining them as essential 
elements of an eternally repeating universe.  
The cyclic model is also inflexible with regard to its 
prediction of no long-wavelength  gravitational waves.

Inflationary cosmology offers no information about the cosmological
constant problem.
The cyclic model provides a fascinating new twist.
Historically, the problem is  
assumed to mean that one must
explain why the vacuum energy of the ground state is zero.  
In the cyclic model, the vacuum energy  of the true ground state is 
not zero.  It is negative and its magnitude is large, as is obvious
from Fig.~1.
However, we have shown that the Universe does not reach the true 
ground state. Instead, it 
hovers above the ground state
from cycle to cycle,
bouncing from one side of the potential well
to the other but spending most time on the positive energy side.

Reviewing the overall scenario and its implications, what is most 
remarkable is that the cyclic model can differ so much from the standard 
picture in terms of the origin of space and time and the sequence of
cosmic events that lead to our current universe.  Yet, the model
requires no more assumptions or tunings (and by some measures less)
to match the current observations.  
It appears that we
now have two disparate possibilities: a universe with a definite 
beginning and a universe that is made and remade forever.
The ultimate arbiter will be Nature.  Measurements of gravitational waves
and the properties of dark energy\cite{other} can provide decisive
ways to discriminate between the two pictures observationally.

\begin{scilastnote}
\item We thank M. Bucher, R. Durrer,
S. Gratton,
J. Khoury, B.A. Ovrut, J. Ostriker, P.J.E. Peebles, A. Polyakov, M. Rees,
N. Seiberg, D. Spergel, A. Tolley, T. Wiseman and E. Witten for useful 
conversations. 
We thank L. Rocher for pointing out historical references.
    This work was supported in part by
  US Department of Energy grant
DE-FG02-91ER40671 (PJS) and by PPARC-UK (NT).
\end{scilastnote}

\end{document}